\documentclass[10pt,journal,compsoc]{IEEEtran}
  \usepackage{array,booktabs,multirow}

\ifCLASSOPTIONcompsoc
  \usepackage[nocompress]{cite}
\else
  \usepackage{cite}
\fi
\ifCLASSINFOpdf
   \usepackage[pdftex]{graphicx}
   \graphicspath{{../pic/}}

   \DeclareGraphicsExtensions{.pdf,.jpeg,.png}
\else
\fi
\usepackage{url}

\usepackage{xcolor}

\begin{document}
%
\title{A Framework for Evaluating Dashboards in Healthcare}
%
%
%
%

\author{Mengdie Zhuang, David Concannon and Ed Manley%
\IEEEcompsocitemizethanks{\IEEEcompsocthanksitem M. Zhuang is affiliated to the Information School, University of Sheffield, Sheffield, UK and the Centre for Advanced Spatial Analysis, University College London, London, UK\protect\\
E-mail: m.zhuang@sheffield.ac.uk.
\IEEEcompsocthanksitem D. Concannon is affiliated to the Centre for Advanced Spatial Analysis, University College London, London, UK.\protect\\
E-mail: david.concannon.14@ucl.ac.uk. 
\IEEEcompsocthanksitem E. Manley is affiliated to the School of Geography, University of Leeds, Leeds, UK  and the Alan Turing Institute for Data Science and Artificial Intelligence, London, UK \protect\\
E-mail: E.J.Manley@leeds.ac.uk.
}

\thanks{(Corresponding author: Mengdie Zhuang.)}

\thanks{This work was supported by i-sense EPSRC IRC in Agile Early Warning Sensing Systems for Infectious Diseases and Antimicrobial Resistance under Grant EP/R00529X/1.}

}
\markboth{Manuscript submitted to IEEE Transactions on Visualization and Computer Graphics}%
{Shell \MakeLowercase{\textit{et al.}}: Bare Demo of IEEEtran.cls for Computer Society Journals}
%



\IEEEtitleabstractindextext{%
\begin{abstract}
In the era of ‘information overload’, effective information provision is essential for enabling rapid response and critical decision making. In making sense of diverse information sources, dashboards have become an indispensable tool, providing fast, effective, adaptable, and personalized access to information for professionals and the general public alike. 
However, these objectives place heavy requirements on dashboards as information systems in usability and effective design. Understanding these issues is challenging given the absence of consistent and comprehensive approaches to dashboard evaluation.
In this paper we systematically review literature on dashboard implementation in healthcare, where dashboards have been employed widely, and where there is widespread interest for improving the current state of the art, and subsequently analyse approaches taken towards evaluation. 
We draw upon consolidated dashboard literature and our own observations to introduce a general definition of dashboards which is more relevant to current trends, together with seven evaluation scenarios - task performance, behaviour change, interaction workflow,  perceived engagement, potential utility, algorithm performance and system implementation. These scenarios distinguish different evaluation purposes which we illustrate through measurements, example studies, and common challenges in evaluation study design. We provide a breakdown of each evaluation scenario, and highlight some of the more subtle questions. We demonstrate the use of the proposed framework by a design study guided by this framework. 
We conclude by comparing this framework with existing literature,  outlining a number of active discussion points and a set of dashboard evaluation best practices for the academic, clinical and software development communities alike. 
\end{abstract}

}

\maketitle

\IEEEdisplaynontitleabstractindextext

%
\IEEEpeerreviewmaketitle

\IEEEraisesectionheading{\section{Introduction}\label{sec:introduction}}

%
%
%
%
\IEEEPARstart{D}{igital} data dashboards are widely employed in modern life, serving as essential tools for performance management to practitioners in a variety of fields. While on the exterior they perform the function of an information access system, providing users with information on critical markers and tracking trends, internally they represent complex systems which interact with or incorporate data storage architectures, state-of-the-art algorithms for query management, information retrieval and visualization, as well as a suite of user oriented features which provide flexibility through personalization and adaptation to various professional contexts.

The design of such systems has received much attention from the research community, however, due to their variety in scope and implementation they have been under-represented in the system evaluation literature \cite{Sarikaya2018}. Indeed, a comprehensive framework for determining when a dashboard fulfills its goals, attains its potential, satisfies users, is robust, scalable and efficient is missing from the literature, hampering dashboard design. 

In the field of healthcare, dashboards have been widely researched, implemented and used for a variety of purposes \cite{dowding2015dashboards}. The data displayed is often complex, for example, individual Electronic Health Records (EHRs) containing large numbers of features with longitudinal changes\cite{west2015innovative}, or disease outbreaks with spatial trends\cite{Harris2018}, or viral genome sequence data requiring additional exploration functions to extract viral evolution insights.  Moreover, the design complexity of healthcare dashboards is leveled up due to the wide spectrum of audiences, ranging from domain experts to the general public (e.g.\cite{Martinez-Millana2019}), together with an extensive variety of use cases, spanning from individual clinical decision making \cite{Dagliati2018}, patient self-monitoring \cite{Soh2019}, and supporting managerial decisions at organisational level\cite{Mahendrawathi2010}. At the same time, there is a strong motivation to develop  efficient healthcare dashboards due to their wide use cases and impact.  Having a well designed health dashboard is not only critical to patients' wellbeing (e.g. \cite{Martinez2018}), essential to keeping key administrative processes in hospitals and healthcare research institutes running on track (e.g.\cite{Mahendrawathi2010,el2011novel}), but also vital for improving the  health literacy of the general public helping them make effective health decisions (e.g. \cite{dong2020interactive}).  Due to the design complexity and interests presented both by the research community, healthcare sectors and society \cite{dowding2015dashboards,Waller2019}, healthcare has seen some of the most notable and refined dashboard models and prototypes being implemented in an effort to attain these goals. The field of healthcare presents most of the theoretical diversity in dashboard types and goals while offering a self-contained window into dashboard design and evaluation. Therefore, in this paper we focus on dashboards in the healthcare domain. 

In this article, we first conduct an integrative review of dashboard evaluations in the health domain grounded in evaluation frameworks of visualization \cite{Lam2011}.  Then we provide a use case of this framework through a report on applying the evaluation scenarios in validating the design of a Coronavirus disease (COVID-19) dashboard iteratively. 
Through our proposed framework, dashboards are cast into the broader context of Human-Computer Interaction (HCI) and Information System Design. Throughout the material and the subsequent evaluation scenarios, we strive to present the dashboard's contribution to solving information needs in a holistic manner, often placing emphasis on interactions that occur outside the dashboard's design scope. This departure from evaluation according to intended design is a novelty in our framework, and encourages a focus on the impact of dashboards not only on its intended user groups but related sectors of the public or community indirectly. We refer to this perspective as \textit{evaluating user outcomes}, and it is an important element in judging the dashboard's overall impact.

To summarize, we make the following contributions:

\begin{itemize}
    
    \item Introduction of a new evaluation framework that contains seven dashboard evaluation scenarios, which extends and refines \cite{Lam2011}, under three evaluation themes. In a rapidly changing field, where novel use-cases and tasks are continuously being developed such a classification is therefore advantageous, as it eschews a major element of variability in dashboard design and focuses on shared design paradigms only. 

    \item Review a diversity of evaluation measures used in practice, and challenges associated with application in the context of  healthcare. 
    
    \item Shift focus from evaluation focused on rigid design paradigms to also encompass real-world outcomes in interacting with broad user groups.
    
    \item Demonstrate using the proposed framework to guide dashboard design through a case study.

\end{itemize}

\section{Related Work }
Several sources in the dashboard literature (e.g. \cite{wexler2017big,few2006information,eckerson2010performance,yigitbasioglu2012review}) highlight the dashboard's visual display and its design purpose as common defining elements. 
Few \cite{few2006information} defines a dashboard as a single screen visual display presenting information needed to achieve a specific purpose, which requires a timely response.  Yigitbasioglu et al.\cite{yigitbasioglu2012review} extended Few's definition by detailing the purposes of a dashboard, in which a dashboard displays ``\textit{the most important information to achieve one or several individual and/or organizational objectives, allowing the users to identify, explore and communicate problem areas that need corrective action}”.  Wexler et al. \cite{wexler2017big} provide a more general purpose compared to \cite{yigitbasioglu2012review}, describing a dashboard as being  ``\textit{... used to monitor conditions and/or facilitate understanding}''. Sarikaya et al.'s review of the design space of dashboards from either the visual genre or function genre  \cite{Sarikaya2018} highlights the lack of consensus on the dashboard's definition. 

Although some evaluation criteria have been applied in practice as evidence of effectiveness (e.g. fitness of the information displayed \cite{eckerson2010performance}, improvements in task performance \cite{Waller2019}, qualitative feedback on user satisfaction and how the dashboard is used \cite{schwendimann2016perceiving}), there is a lack of systematic discussion on the evaluation approach dedicated to dashboards \cite{verbert2014learning}.  The closest related approach lies in evaluating information visualization, which has been discussed from the perspective of a break-down list of desired quality \cite{Zhu2007}, design processes \cite{Munzner2009}, or evaluation scenarios \cite{Lam2011}. However, as dashboards are incorporated in various socio-organisational contexts, they take on the role of more than just a visual display, or interface.  As stated by El-Turabi et al. \cite{el2011novel}, dashboard design within health research systems ``\textit{requires a full understanding of the operation of the organisation and the collaboration of its employees}''.

Uncertainty around the definition and associated evaluation approaches has not prevented dashboards becoming critical in serving timely and complex information to the user. Nowhere is this more the case than in healthcare, where growing interests of deploying dashboards have coincided with the increased use of EHRs and publicly available health information. Traditionally, visualizations, including scatter plots and time-series graphs, have been widely built into dashboards to help inform clinical decisions about individual patients\cite{powsner1994graphical,Plaisant1996} or managerial decisions (e.g. logistics) for organisations such as hospitals\cite{Mahendrawathi2010}. More recently, in response to COVID-19, dashboards containing geographic maps have been used to collate real-time statistics of outbreaks, which keeps the public informed, and serves as a shared data collection to support decision-making for policy makers and researchers\cite{Harris2018,dong2020interactive}. The increased number of use cases of different dashboards is accompanied by a diversity in user groups, for example, to tailor treatment plan for individual clinicians \cite{Patel2018,Peterson2014}; to monitor key health indicators for patients and carers\cite{Soh2019}; to learn a health related concept for the public\cite{dong2020interactive}; or to monitor performance at a team \cite{Martinez-Millana2019} or organisation level\cite{Mahendrawathi2010}.   Subsequently, this complexity associated with healthcare dashboards audience and use cases, which is also observed in \cite{Sarikaya2018} for dashboards in other domains, leads to the need for dedicated consideration when selecting appropriate effectiveness measurement and evaluation methods. A number of reviews have discussed dashboards and visualizations in the health domain.  West et al.'s \cite{west2015innovative} review of the use of information visualization in EHR in the literature concluded that while most of the studies surveyed the importance of the growing amount of clinical data they found there is little focus on using innovative visualization techniques that lend themselves to the large complex datasets available electronically. Dowdling et al.'s \cite{dowding2015dashboards} review of dashboards for improving patient care concluded that there was some evidence that the use of dashboards improved patient outcomes, although it is unclear what dashboard characteristics lie at the root of these improvements. A more systematic evaluation framework could serve to shed light on these connections and provide an environment for researchers to conduct more conclusive experiments on dashboard design and implementation.

\section{Methodology}
\label{sec-method}

\subsection{Definitions}
\label{sec-codes}

As described above, dashboards are demarcated by their visual design and purpose \cite{Sarikaya2018,yigitbasioglu2012review}. Considering the growing analytical functions dashboards facilitate, they continue to progress beyond the capabilities of a mere data display.
Dashboards are examples of \emph{visual} information systems (see discussions on definition of information system in \cite{valacich2009,rainer2020introduction}), by which we mean information  systems which primarily employ a \emph{visual medium} for communicating information. As such, their purpose is accessing and understanding data \textit{relevant for a concrete problem}, and their means of achieving this purpose is inherently visual. In this respect, they differ from \emph{information visualization systems}, whose purpose is \textit{designing, building or refining visualizations of data} to reinforce human cognition (often in the form of editors, or software meant to serve this purpose, which can themselves be used to create dashboards, e.g. Tableau), whereas dashboards eschew this goal completely: dashboards instead serve as ecosystems for integrating visulazations, with the sole purpose being accessing and exposing data relevant to their task.

Dashboards come equipped with a graphical user interface (GUI) which presents \textit{an interaction and information access channel} to the user. At any one time, the GUI displays a fixed view of the information available to the user which may be a subset, reduction, aggregation etc. of the  data available to the dashboard. The primary characteristic of the dashboard is that it varies the information in its view according to changes in the data (e.g. tracked over time, space or other variable parameters) or as a response to user interactions (e.g. clicks). Therefore, a dashboard needn't be physically interactive, but then it must reflect the changes in the data and update the view on its own; conversely, it needn't display changing data, but it must present the data with variations as requested by users, to tailor to their information needs. All dashboards encountered in our review sit on a scale between these two extremes. 
Summarizing the above, we give a more general, broadly encompassing definition of dashboard below and position the rest of this article with respect to it:

A \textbf{dashboard} is a \emph{visual} information system which comprises at least one visual figure and a store of data which the GUI exposes, which is designed and built with the purpose of fulfilling a precise information need.

Part of the definition is the dashboard's purpose in fulfilling a particular information need. In order for designers to guide the dashboard towards its purpose, a particular type of task or set of tasks is designed and engineered into the dashboard. 
Such tasks are developed in relation to context-driven problems that the dashboard is intended to tackle, and we refer to them as the dashboard's \emph{intended} tasks.

Upon deployment in their respective ecosystems, evaluation by intended task quickly becomes stale for most dashboards, especially those targeting critical subject matter such as healthcare. In the following framework we pay attention not just to the evaluation by intended task, but collect numerous examples of positive or unsatisfactory outcomes from the use of the dashboard in the wild. Though less structured, this form of evaluation still holds value and informs designers on the multilateral character of dashboards in the way they communicate data. As such we refer to \textit{user outcomes} as the final end-point in the evaluation pipeline where we observe positive or negative repercussions of the interaction with the dashboard by diverse user groups with goals outside the immediate scope of the dashboard's design. We consider this type of empirical evaluation valuable and collect relevant examples from published studies.

\subsection{Review Method}
\label{sec-reviewmethod}

\begin{figure*}
    \centering
    \includegraphics[width=2\columnwidth]{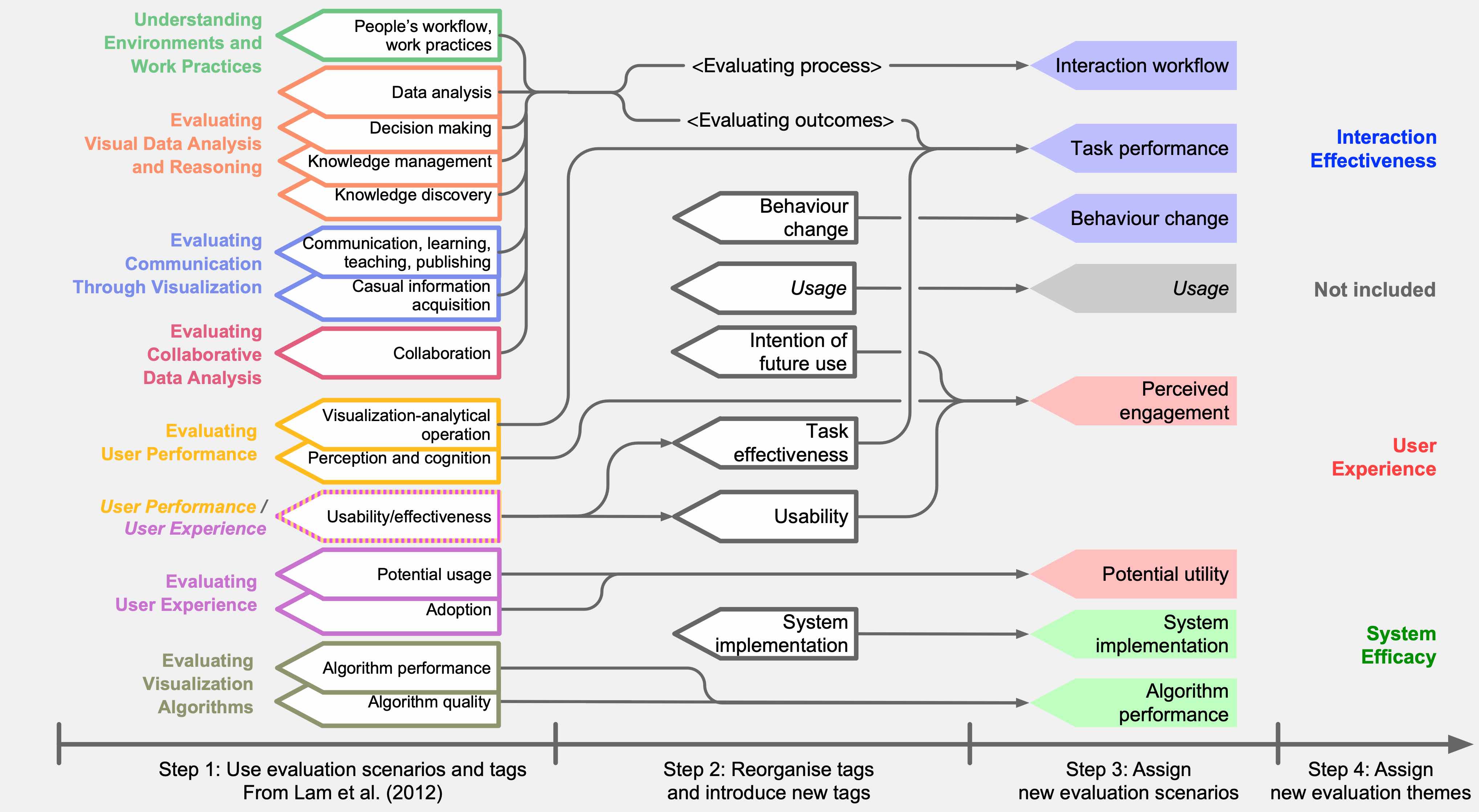}
    \caption{The development of evaluation scenarios in this study. The coding process started from using the collection of 15 tags and seven evaluation scenarios proposed by Lam et al.\cite{Lam2011} (step 1). Each of the 81 papers was coded twice. Three discussion sessions among all the co-authors were conducted during the review, and the collection of tags was reorganised, including adding new tags and grouping similar ones,  during the discussion (step 2), which further formed the group of evaluation scenarios (step 3). Another discussion was held to form the evaluation themes (step 4).}
    \label{fig:tagsprocess}
\end{figure*}

The relevant literature was systematically searched and synthesized following five stages \cite{whittemore2005integrative} (problem identification, literature search, data evaluation, data analysis, presentation) to conducting an integrative review in order to incorporate studies with diverse methodologies, and to critically analyze the literature and bring a new understanding of dashboard evaluation. 

\textbf{Problem identification}: This review scrutinizes the current dashboard evaluation literature, aiming to single out the
most prevalent issues with classifying evaluation approaches and establishing consistent evaluation methodology. We aim to address this problem in the context of healthcare where we have found it prevalent and where the need for clarity in goal and scope is essential due to the time-sensitive and life critical implications that these systems carry.
An integrative review of the related literature was able to reveal: (i) the main types of dashboard evaluation in the health domain; (ii) the criteria and common measurements that have been used to evaluate dashboards in the health domain; (iii) the challenges of applying such evaluations. (iv) given (i)–(iii), the best practices to evaluate dashboards. The focus of this paper is on measurements, criteria and how they have been used in evaluating dashboards in healthcare. Other aspects related to evaluating dashboards are not reported.

\textbf{Literature search}: To get a comprehensive overview of how dashboards are evaluated in healthcare, IEEE Scope, ACM Digital Library, Google Scholar, and PubMed were first searched. A combination of the following terms (including terms obtained through affixation) was used to search in titles, abstract and keywords: dashboard, evaluation, measure, and health. The search was conducted on the 4th December 2019. At the same time, we adapted the forward and backward snowballing approach \cite{Wohlin2014} in order to follow the references of identified literature and the works which cited them. The initial search resulted in a selection of 260 articles.

\textbf{Data evaluation}: A paper was classified as relevant if all of these conditions apply: i) it addresses a dashboard or a system contains a dashboard, ii) it belongs to the domain of healthcare or has the objectives which benefit the health context, iii) it describes any evaluation, assessment or measurement of the quality of the dashboard or system. A paper with mentioned terms was classified as irrelevant if: i) the dashboard is used as an evaluation of a separate project, ii) the content lies outside the human health domain (e.g. dashboards that track the `health' of a non-health related project). Study protocols were excluded in the review. 

Abstracts of the initial selection (260 papers) were first screened by the first author against the relevance criteria, yielding 134 relevant papers and 126 irrelevant papers. Three coders (all co-authors of this paper) participated in the assessment of the full text of the 134 relevant paper in order to further remove papers against the relevance criteria mentioned in the previous step, or the papers which only have the abstract available, or the papers describing the same dashboard, which yielded a final corpus of 81 publications.

\textbf{Data analysis}: The 81 publications contain details of evaluations of 82 healthcare dashboards.  
For coding dashboard evaluation, we use a hybrid approach that combines deductive and inductive coding following the steps in \cite{fereday2006demonstrating}. A deductive approach better places the analysis in a broad context and connects our results with pre-established theories in other fields, while an inductive approach allows the inclusion of new themes emerging during the data analysis. 

Fig. \ref{fig:tagsprocess} illustrates the four development steps of our final codes compared to the initial schema. We based our initial coding scheme on the seven scenarios and associated tags proposed by Lam et al. \cite{Lam2011} for evaluation of generic information visualization (step 1 in Fig. \ref{fig:tagsprocess}). The authors of this paper held three discussions during the review regarding the validity of the initial code scheme and the potential need for extension.  Due to the increasingly diverse functionality of dashboards \cite{Sarikaya2018}, we reorganised the eight tags in Understanding Environment and Work Practice, Evaluating Visual Data Analysis and Reasoning, Evaluating Communication through Visualization, and Evaluating Collaborative Data Analysis scenarios in \cite{Lam2011} by their evaluation focus, either on the quality of interaction process or outcome.  In addition, we broke up one tag ‘Usability and effectiveness’, which belongs to two scenarios Evaluating User Performance and Evaluating User Experience, into task effectiveness and usability. We further extended the initial code schema by adding four codes, behaviour change, usage, intention of future use and system implementation based on observations from the literature. Behaviour change is selected from the psychological research and widely used in public health \cite{michie2012theories}; Intention of future use is selected from HCI \cite{OBrien2008}; System Implementation is selected from the software engineering domain \cite{ISOIEC25010}. We decided to not include one tag, namely usage, namely the descriptive statistics alone of the dashboard used (e.g. the number of users viewed this dashboard, average time spent on the dashboard) without any benchmark. The popularity of dashboards is a result of many factors and is not evidence of the dashboard's quality. Step 1 and 2 in Fig. \ref{fig:tagsprocess} illustrate all the changes we made on the tags. 

We further group or pass the remaining and newly introduced tags into scenarios (Step 3 in Fig. \ref{fig:tagsprocess}). As visualization analytical operation and task effectiveness also examines the outcomes of using the dashboard, we group them with the evaluating outcomes category into one scenario, Task Performance. Three tags, intention of future use, perception and cognition, and usability use mainly self-reported subjective measures to evaluate existing functions or aesthetics of the dashboard, therefore they are grouped into the Perceived Engagement scenario. Potential utility instead focuses on possible usage or functions that are not included in the current tested version, therefore left as an individual scenario. Algorithm related tags stayed the same. Seven scenarios were generated belonging to three general themes: Interaction effectiveness, User experience, and System efficacy (Step 4 in Fig. \ref{fig:tagsprocess}). The scenarios belonging to the interaction effectiveness themes are: Interaction workflow, Task Performance, and Behaviour Change. The scenarios under the user experience theme are: Perceived Engagement and Potential Utility. The scenarios under the system efficacy theme are System implementation and Algorithm Performance.

\textbf{Presentation}: The full set of papers with their codes can be found at: https://tinyurl.com/dashboardscenario. 
Evaluation scenarios are discussed in section \ref{sec-scenario}, as are relevant measurements and challenges from the literature.

\section{Evaluation Scenarios}
\label{sec-scenario}

Evaluation methods and objectives typically address various key evaluation metrics which are specific for each domain. Generalizing across multiple dashboards, evaluation metrics coalesce under particular themes specific to different facets of the interaction. For illustrative purposes, and for consistency with  Lam et al \cite{Lam2011} we adopt \emph{evaluation scenario} as an umbrella term for all methods and objectives that address a single, particular aspect of the interaction between user and dashboard. These are not specific to healthcare, but in our analysis, the methods and criteria exemplified will be.

We extracted seven evaluation scenarios representing the categories of evaluations we found in our literature review. We provide a definition of each scenario, identify the common evaluation questions, measurements and challenges with examples in healthcare. The seven evaluation scenarios may be more or less appropriate according to the purpose and implementation details of different healthcare dashboards, examples of which are given in each scenario. 

We group these scenarios into three themes: \textit{interaction effectiveness}, \textit{user experience}, and \textit{system efficacy} (step 4 in Fig. \ref{fig:tagsprocess}). In the interaction effectiveness group, the goal of the evaluation is to measure how effective the dashboard is while the user interacts with it, focusing on how the interaction develops, the effects of interaction on a task, or in the long run between the user and the dashboard. As invoking interaction between users or enhancing data exploration, organising data for the user(s) is the main motivation of using visualizations, most of the visualization evaluation scenarios \cite{Lam2011} are grouped into this theme. In the user experience group, the evaluation only focuses on users' subjective feedback in terms of usability issues, extra functionality to include and the intent to engage with the system in the future. In the system efficacy group, the main goal is to understand whether the system contains an accurate algorithm or has stable outcomes. We also care about the quality of the implementation of the dashboard, for example whether the data presented, and the functions included suffice for the intended task. Usually dashboard evaluation contains more than one type of the three in the process of design or post-deployment.

\subsection{Task Performance (TP)}
\label{effectiveness-1}
\textit{How does the use of the dashboard influence expected task outcomes in the dashboard's intended task?}

\subsubsection{TP: Evaluation Aim}

Evaluating task performance aims to \textit{assess the effectiveness of the dashboard with respect to the performance of a particular task(s) across multiple users}. Examining how a dashboard facilitates task performance is the second most used evaluation scenario in the literature review (43 out of 82, 52.44\%). This scenario is intrinsic to a coupling of dashboard 
and task. As discussed in Section \ref{sec-codes}, such dashboards comes
engineered with an intended task. When speaking of the evaluation of a particular dashboard with respect to task performance we refer to its evaluation with respect to this task or set of tasks. Alternatively, through real-world usage or through serendipitous exposure outside the designer's intent, the dashboard may be subjected to use for secondary tasks. These usually appear as a result of interaction between the dashboard system and the polarized needs of particular user groups. One may attempt to assess the dashboard with respect to such a task as well. 
We dedicate an evaluation scenario \textit{potential utility} (Section \ref{experience-2}) to identifying these secondary tasks.

\subsubsection{TP: Evaluation Criteria}

Evaluating task performance is a \emph{post factum} type of evaluation in which researchers employ quantitative or qualitative measures of the task outcomes (typically averaged across many users to remove users as a source of variance).
When discussing \textit{improvements in task performance} it is these refined quantitative metrics of the task outcome that researchers most often refer to, for example completing a designed task in shorter time, with fewer total actions, higher accuracy, and higher occurrence of desired actions.
Benchmarks or other adequate means of comparison still need to be provided in order to ground the evaluation, but this is in general the most straightforward evaluation context and also the second most widely considered. 
 
\subsubsection{TP: Measures and Examples}

Typical task outcomes include the percentage or accuracy of task completion \cite{Concannon2019,Martinez2018}, time to completion \cite{Dickson2017,Azad2016}, time to make decisions \cite{Dolan2013}, effective actions triggered (e.g. clinicians' response to alerted high-risk medication \cite{Patel2018}, types of clinical actions taken with and without the dashboard \cite{Dagliati2018}), or quality indicators (e.g. drug to drug interaction alerts \cite{Simpao2015}). The measurements do not have to be linked with the direct operation of the dashboard, as the dashboard can issue alerts or reminders of external events or the degradation of certain KPIs that are handled or dealt with by the user externally 
(e.g. number of visits of chronically ill patients to clinics \cite{Peterson2014}). 

However, such outcomes might not be easily measurable for dashboards with a tracking purpose or those which rely on the users to explore the content (e.g. teaching, knowledge discovery, and tracking workouts), as a successful use session may have higher duration in this context and the performance is laborious to quantify or benchmark. Researchers have proposed dealing with this issue via two approaches. 

The first is relevant when the difficulty in measurement stems from the infeasibility of collecting
reliable data. 
The solution in this case is to add constraints to the evaluation task design, in order to facilitate the observation of the desired outcome. An example would be to shorten the task session length or instruct the participant to focus on one sub-task which they would usually do in a casual situation as seen in Bernard et al. \cite{Bernard2019}, in which a dashboard that helps the user to extract longitudinal and cross-cohort patterns from patients' medical history is evaluated by measuring the observations users made in a 20 minute session while being clearly instructed to compare the given patients. 

A second approach involves lowering the evaluation threshold, by testing the minimal necessary functionality of the dashboard. In practice, this involves testing not task performance for the intended task, but the performances of the minimally required executions for the intended tasks, a less elucidating measure, but nevertheless more straightforwardly computable (circumventing issues of user variance and objective measurability). For example, in a diabetes patient self-management and learning dashboard \cite{Martinez2018}, researchers selected three sub-tasks, which are the basic required tasks for users to use this dashboard, identifying recent hemoglobin, messaging a doctor, and setting a reminder for a clinical visit. The results provide evidence that the participants do not experience difficulties completing these three sub-tasks, therefore the tool has the potential to at least facilitate these minimum required task for self-management. However, the result is insufficient in supporting this tool for self-management and learning purposes as the outcomes are not measured.

\subsubsection{TP: Evaluation Challenges}
\label{tpchallenge}

\textbf{Selection of tasks and sub-tasks}. Choosing a task is necessary for this evaluation.  Although this statement is trivial for dashboards with a single intended task, (e.g. the door-to-balloon time for stroke treatment in the emergency department \cite{Dickson2017}), it needs more consideration for dashboards with a tracking purpose or those centred on exploration. 
In either of these cases, focusing on a specific task is not always straightforward and checking through all possible use cases is not feasible in practice. Researchers typically extract a couple of essential or frequent subtasks from the primary one, and apply subsequent evaluation criteria to these. For example, 
Pickering er al. \cite{Pickering2015} examined a data management dashboard, by timing how long clinicians spent on collecting the most used clinical data. Inheriting the key considerations in evaluation from the dashboard design is crucial. Concannon et al. \cite{Concannon2019} introduces a design which tackles visualization literacy, and creates the information extraction tasks from the dashboard that involve participants representing different visualization literacy groups.

\noindent\textbf{Performance aggregation}. As mentioned, in order to evaluate tracking or exploration focused dashboards, researchers select several sub-tasks that are either essential sub-tasks or frequent sub-tasks with respect to the intended one.  Researchers ultimately need to translate evaluation of individual sub-tasks into an evaluation of the primary task which is linked to the design and nature of each particular dashboard. Such a process requires the aggregation of scores and measures, a procedure for which there is no theoretical prior. In practice,
aggregating the results requires domain specific knowledge and, most typically forming a priority list of criteria or a weight. 
For example, Azad \cite{Azad2016} introduced a system that collects spine surgery outcomes and displays these data with clinical records to the clinicians, and evaluated the survey data capture rate, which is the primary outcome of the data collection sub-tasks as well as the average visit time of patients, which is the overall expected outcome. Patel et al. \cite{Patel2018} also picked multiple performance measures, including time spent before making the decisions, which is the efficiency, and the drugs picked, which infers the accuracy.

\subsection{Behaviour Change (BC)}
\textit{How does the use of the dashboard induce long lasting behaviour changes in user groups?}

\subsubsection{BC: Evaluation Aim}
Behaviour change (reported in five out of 82 evaluations, 6.1\%), a new scenario included for dashboard evaluation, aims to \textit{assess the dashboard's ability to induce positive long term influences on users' behaviour}. 
The dashboard provides a particular information access experience  which impacts the consequences of our actions and gives us agency in our environments. Furthermore the interaction with the dashboard serves to shape the user's workflow and awareness, and induce certain behaviours and needs as a consequence. This scenario aims to evaluate the extent to which such interactions can have a long-lasting impact on users irrespective of the behaviours determined at session level interaction. 

This scenario is different from the evaluation of \textit{task performance} in which the dashboard directly contributes to the performance evaluation. Behaviour change is typically independent of the direct or session-level interaction with the dashboard, is driven primarily by awareness or interest on the part of the user, and one of its characteristics is its staying-power, describing a long term influence. Behaviour changes are also often associated with dashboards whose purpose is to raise awareness, foster interest toward certain issues or mark the appearance or presence of certain patterns, or facilitate a change in organisational process. 
Inducing positive behaviour change in users and patients represents one of the most highly sought after features of dashboards in healthcare. 
However, while behaviour change is stipulated as the ultimate motivation of implementing certain dashboards, evaluation of behaviour change for the very same systems often falls short of expectations.

\subsubsection{BC: Evaluation Criteria}

Evaluating behaviour change is potentially one of the most difficult tasks for an investigator, due to the time required to observe such changes and the lack of reliability in reported measures (often we have to rely on the self-awareness of subjects themselves to make the measurements available). 
In general, it is a challenge to describe types of behaviour as either positive or negative from a psychological perspective. In the field of healthcare\cite{michie2012theories}, however, \textit{positive behaviour changes} can more easily be defined: for patients and the general public they are those which induce a heightened awareness of their state of health and determine them to improve it, whereas for clinicians and medical staff it involves a deeper understanding of the status and needs of their healthcare ecosystem, and an updated, more effective approach to the access and analysis of information. 

\subsubsection{BC: Measures and Examples}
Any change in actions, patterns or opinions of a user or user group, or changes in organisational process \emph{can} be attributed to behaviour change. How this attribution is made is the responsibility of the researchers conducting each study. What is common across different scenarios is the data collection methodology for tracking such changes. Examples include tracking changes in awareness of a particular topic or issue, changes in professional behaviour or even changes in lifestyle, changes in organisational or recommended procedure of completing a task (e.g. prescribing high-risk drugs).
Alternatively, one can observe the ostensible consequences of these changes, however, this creates even more uncertainty around attribution.

The measurements of behaviour change can be collected from users directly (e.g. self-reported knowledge gain \cite{Zahanova2017}), or as an observation from a third party (e.g. collecting community health workers' home visit time from visited households \cite{Whidden2018}), or as the result of such behaviour changes (e.g. measuring the impact of clinicians' awareness of certain clinical guidelines or current situation by clinical outcomes, such as patient re-admissions \cite{Park2019}). 
A variety of consequences are usually collected by the researcher to demonstrate a change in behaviour. For example, Hull et al. \cite{Hull2014} evaluate the monthly clinical targets tracking dashboard that aims at raised clinicians' awareness of these targets using influenza immunisation rate and care plan completion rate in the community served by the clinics. They assume these rates are affected by clinicians' behaviour change. In addition, Touray et al. tested a dashboard for tracking the vaccination teams' settlement coverage \cite{Touray2016} using measures including the geographical coverage of settlements and the number of missed settlements and, as a result, workers started visiting a wider range of settlements.

\subsubsection{BC: Evaluation Challenges}

\textbf{Establishing causation.} Linking the occurrence of a change in behaviour to positive features of the dashboard  remains a challenge. As causation is usually difficult to determine, most studies focus on determining correlation instead. This issue is more relevant in evaluating behaviour change due to the uncertainty of when such changes will happen,
and the long time required for effects to develop or stabilize.
Furthermore, behaviour change can be the result of a combination of factors, and the long duration of inducing behaviour changes makes it harder to disentangle the relationships between them. 
Overall these pitfalls point to a rugged research landscape riddled with the danger of establishing spurious statistical implications. Although methods have been developed to tackle causality in other problems in healthcare (e.g. \cite{greenland2002overview}), we have not observed any dashboard evaluation studies employing them to establish causality empirically.

\subsection{Interaction Workflow (IW)}
\label{effectiveness-3}

\textit{How intuitive is the dashboard to use when executing common interaction patterns and analysis tasks?}

\subsubsection{IW: Evaluation Aim}
Evaluating interactive workflow aims to \textit{assess how users interact with a dashboard from the point of view information seeking, communication and decision making efficiency for the intended task and context}.  Although being the long focus of the Human Computer Interaction (HCI) and the Information Seeking  communities (e.g. \cite{Case2016}), remarkably only a few dashboard evaluations looked into interactive workflow (seven out of 82, 8.54\%). As most researchers attempt to sketch the optimal interaction workflow through discussion with domain experts in the dashboard design stage, they employed other evaluation scenarios (e.g. \textit{task performance}) as validation.  The typical workflow for dashboard usage is similar to that of a general information system, in that information seeking steps alternate with micro-decision making steps in active - reactive phases (e.g. berry picking model \cite{Bates1989} in information seeking and retrieval, Norman's interaction model \cite{Norman1986} in HCI, NOVIS model \cite{Lee2015} in visual sense making). Thus, irrespective of the available information in the data and the desired information goal, there can be many avenues for the user to attempt a task. This evaluation scenario focuses on determining how laborious or strenuous the average interaction workflow is from this perspective - is the interaction natural and intuitive to the user or convoluted and opaque? A simple yet effective information workflow is the hallmark of a good dashboard. 

\subsubsection{IW: Evaluation Criteria}

What constitutes a positive interaction workflow and methods for how to compare different technologies according to this criterion have proven to be central topics in the HCI community since its inception. However, these questions have yet to receive a uniform answer, with many methods and guidelines being developed to suit case-studies (e.g. \cite{reimann2009time,sanderson1994exploratory}).  In the case of dashboards in a critical domain such as healthcare, evaluation used to be primarily outcome-driven (e.g. \cite{Dickson2017}), however recent trends put more focus on clinical practitioners as well as patients as users. In order to induce a \emph{positive interaction workflow}, researchers attempt to streamline the interaction in terms of  minimizing time spent, cognitive load and total number of user-issued actions as a cost for information acquisition as well as 
increasing the confidence in the interaction.

\subsubsection{IW: Measures and Examples}
The interaction of users with information systems is complex and occurs across multiple channels.  The commonality of the theoretical models mentioned above describing the user interacting with a system is that the interaction exists on two separate levels, physical and cognitive. The physical interactions are represented by the actions (e.g. mouse clicks, keystrokes) the user issued, and the cognitive interactions are represented by the information wrapped in the dashboard noticed or examined or paid attention to by the user and the decisions made while examining the data.  Typically, interaction data are collected through automatic behaviour recording \cite{Koopman2011}, observation \cite{Ni2019}, interview \cite{Jeffries2018}, focus group, and think aloud \cite{Dagliati2018}. 

The key questions researchers pose are concerned with the data examined, the nature of the user's actions, and the order in which they occur (e.g. the order of the screens scrolled through
\cite{Koopman2011}). The associated costs (e.g.  time invested, cognitive load, decision points) and the information extracted (e.g. number reading, the complete decision tree \cite{Meijers2013}) also factor into the evaluation of interaction workflow.

In other HCI studies, techniques from psychophysiology  (e.g. Electroencephalogram \cite{Cacioppo2007}) have been used to collect information related to the psychological processes associated to the interaction such as motivation, cognition, emotion, learning during the interaction. Several advantages, including being more directly connected to users, more objective, and the ability to capture changes over time \cite{Lopatovska2011}, resulted in their increased popularity in the information retrieval field and in visual search field to track attention and cognitive load (e.g. \cite{Moshfeghi2016,kamienkowski2012fixation}). We haven't noticed any health dashboard studies adapting such techniques, which may be due to the case specific nature of dashboard design so far. However, it is worth noting that they can serve as a method to collect interaction data for dashboard evaluation.

\subsubsection{IW: Evaluation Challenges}

\textbf{The effect of individual variance.} 
For all quality measures which involve the participation of users, the practice of aggregating statistics across users in order to obtain user-independent outcomes which pertain to the quality of the dashboard alone is ubiquitous within evaluation methodology (e.g. \cite{Kelly2009}). Users exhibit variance due to a myriad of characteristics (in the case of dashboards most relevant being prior domain knowledge, visual literacy\cite{Sarikaya2018}) which can bias the analysis of the system if not accounted for. Whereas quantitative measures can be aggregated to compute expected values over a population, categorical distributions such as interaction patterns do not admit a simple definition of expectation. 

\noindent\textbf{Extracting sequential information from interaction.}
A large number of the studies which report interaction workflow evaluation methods, fail to encompass the full complexity of user interaction sequences into their analysis, relying instead on proxies with limited descriptive power (e.g. total time spent interacting with the interface, the data that got attention \cite{Koopman2011}) or merely on the results of a qualitative analysis (e.g. \cite{Dagliati2018}). However, more useful mathematical models  for abstracting interaction sequences (e.g. \cite{Drutsa2015}) are readily available, but require more refined analysis and are still at the fringe of techniques employed by the community.

\subsection{Perceived Engagement (PE)}
\label{experience-1}

\textit{Do users feel engaged in the interaction with the dashboard beyond the immediate task completion utility?}

\subsubsection{PE: Evaluation Aim}
Evaluating perceived engagement aims to \textit{assess the subjective feedback collected directly or indirectly from users, regarding their perception of the user experience}. It is the most deployed scenario in dashboard evaluation as 49 out of the 82 papers (59.76\%) contained at least one perceived engagement evaluation. Providing users with a positive experience is the ultimate goal of developing any type of information system. In particular, the only gateway into evaluating perceived engagement is the feedback obtained directly from users themselves.  Furthermore, the efforts spent on developing valid and reliable perception measurements through a structured developmental process lead to a set of standards and measures that are ready to use, and easy to compare across studies. Perceived engagement appears in almost all dashboard lifecycle stages, such as design, and the validation of implementation.  Measuring perceived engagement is typically obtained from data collected directly from users through questionnaires and interviews. Such methods require participants volunteering their responses to a set of questions or following a set of instructions. Therefore, data collection for evaluating perceived engagement is obtrusive to users’ natural interaction with the system. 

\subsubsection{PE: Evaluation Criteria}

Perceived engagement represents the user's perception, affective capability, mood, emotions and intentions as a product of their interaction with a system \cite{Lalmas2014}. An \textit{engaging experience} emphasises positive aspects of the interaction – that the users not only feel the technology is easy-to-use, are satisfied with the interaction, but also feel a sense of reward from the exchange and therefore want to use the technology longer and more frequently \cite{OBrien2008}.

\subsubsection{PE: Measures and Examples} 

Various dimensions of user perception are employed to assess perceived engagement, and, together with their associated instruments for measurement (e.g. questionnaires, interview framework) they are extensively discussed in the HCI community. These dimensions initially emerged from the perceived usability \cite{Nielsen1994} family of studies, prominent ones including interface aesthetics, satisfaction, perceived difficulty of using the dashboard to complete a task, and perceived usefulness. The usability-related dimensions are measured via a wide range of methods, and a selection of methods tend to be consistently deployed (e.g. questionnaires such as the System Usability Scale (SUS) \cite{Brooke1996} in \cite{Soh2019,Yoo2018,Martinez-Millana2019,Dagliati2018,Ni2019},  Computer System Usability Questionnaire (CSQU) \cite{Lewis1995} in \cite{Martinez2018}, and interviews \cite{Whidden2018}).  

In addition to the usability group, the \textit{acceptance} of a new tool and the \textit{intention} to use in the future are the other two main dimensions that have been assessed. Acceptance of a new tool measures a user's intention to use an information system and subsequent usage behavior subject to the effort the user requires to invest in order to adapt to the tool. Acceptance was investigated both at individual level (e.g. through the unified theory of acceptance and use of technology model \cite{Venkatesh2003} in \cite{Dolan2013}), and at the organisation level (e.g. using theory of organizational readiness for change \cite{Weiner2009} in \cite{Dunn2015}). 

Although using a mixed sets of dimensions will certainly provide better coverage of the user experience, it may prove too laborious for the participants to respond to. In addition, these dimensions are correlated with one another (e.g. user perceived satisfaction, aesthetics and usability \cite{Tractinsky2000}, perceived usefulness and satisfaction \cite{Calisir2004}), as most of them were developed from usability research, which represents an additional shortcoming. 

\subsubsection{PE: Evaluation Challenges}
\textbf{Obtrusive to natural interaction.}
Perceived engagement data is mainly collected using self-reported methods, such as interviews, diaries and questionnaires. Thus it requires user responses to a set of questions or items or following a set of instructions. Such events are obtrusive to a user’s natural interaction, interrupting the flow of the user experience, and making the collection impractical with large instruments (e.g. large number of questions or long guidelines). Therefore, dynamically assessing perceived engagement (e.g. at a certain points in the middle of the session) during the interaction has not yet been conducted extensively.

\subsection{Potential Utility (PU)}
\label{experience-2}

\textit{How much potential does the system have for integrating useful future functions and features?}

\subsubsection{PU: Evaluation Aim}
Evaluating potential utility aims to \textit{assess the dashboard's potential secondary tasks and supplementary functionality in addition to its current design or use.} Evaluation of potential utility should give an indication of possibilities for applications that have not yet materialized but are currently deemed useful or impactful within a professional community. This type of evaluation is variable and subjective, and usually collected though self-reported methods such as interview, think-aloud or questionnaires. The resulting analysis redoubles the need to consider secondary tasks - those not intrinsically designed within the dashboard, but still actively performed by its user groups - or functionalities that support the two kinds of tasks. 
Potential utility is dependent on the dashboard's ability to foster the creation and definition of secondary tasks. Therefore, it is crucial for the dashboard that aid in real-time tracking of data and statistics and decision making to be mindful of potential utility. Six out of 82 dashboard evaluations (7.32\%) contain this scenario. 

Different from the \textit{system implementation} scenario (Section \ref{system-2}), potential utility focuses more on functionalities not purposefully built in at design time for the intended task, but which still surface through use. They are inherent to the dashboard and may not be fully implemented, but there is an indication that the overall system is capable of supporting such a functionality. Potential utility provides directions for improvements of the dashboard, therefore it is widely used in iterative design (e.g. \cite{Martinez2018}).

\subsubsection{PU: Evaluation Criteria}

A\textit{ high potential utility} corresponds to a system that enables easy extension, repurposing and addition of functionality. 
On the other hand, it may be hard to disentangle the evaluation of potential utility from shortcomings of the dashboard's current design, i.e. researchers should take care not to misattribute the dashboard's current lack of essential functionality for the intended tasks to potential future developments.

\subsubsection{PU: Measures and Examples} 
To evaluate potential utility,  data is usually collected through self-reported methods such as interview, and questionnaire. These collection methods make potential utility evaluation usually exist as part of the designed survey, think-aloud study, open discussion or interview that are also employed in collecting data for evaluating perceived engagement (e.g. \cite{Concannon2019}) or system implementation (e.g. \cite{Lenglet2019}).

Potential utilities comprise functions, data presentation, or potential use cases that are not involved in the current version of the dashboard. 
Alert functions for tracking dashboards are universally desirable (e.g. \cite{Koopman2011,Soh2019}), and have been signalled out by several studies. For instance, if users develop a predilect way of interpreting the data, automating such functionality becomes desirable; this turns a dashboard for tracking into an alerting dashboard and, in turn,  reduces users' effort of extracting this information actively. For example, preoperative gastric cancer patients wished to have an alert function for a self-monitoring tool \cite{Soh2019} on usage of an incentive spirometer, which is a medical device used to help patients improve the functioning of their lungs. 

Frequently, studies incorporate some form of potential utility results when participants accidentally mention these features (e.g. in think-aloud study), without being explicitly instructed to do so. We omit these studies for this evaluation type as the evaluation are not purposefully designed to assess potential utility.

\subsubsection{PU: Evaluation Challenges}

\textbf{Effective follow-ups.} 
Assessments of potential utility typically require follow-ups from both the researchers and the dashboard designers in order to achieve full potential. Researchers need to abstract the problem, explicate the required features and put them into domain context. 
Often, potential utility is merely summarily reported, but interpretation and guidance on the part of researchers could provide the key insight to implementing these features or use cases in future updates of the dashboard. For example, many individual features may be proposed by participants which collectively address the same problem, it then rests on researchers to define the scope of the requirement and extract the most appropriate features which address the problem while facilitating a large spectrum of the proposed interaction types.

This challenge is however well addressed in studies which report iterative design and evaluation patterns, as the potential utility evaluation is repeatedly fed back into the design process. 
For example, Martinez et al. \cite{Martinez2018} include the potential utility evaluation in the iterative design sprint, in which they ask users what other features they would like to have access to in the dashboard (e.g. the blood test results of patients-like-me value). These features are then included in the next version for testing until no more new features were suggested by the participants. 

\subsection{Algorithm Performance (AP)}
\label{system-1}

\textit{Does the algorithm have accurate and efficient outputs?}

\subsubsection{AP: Evaluation Aim}

This scenario aims to  \textit{assess whether the algorithms embedded in the dashboard are correctly designed and return reliable information efficiently.} Note that this scenario is relevant for some dashboards with built-in algorithmic components that process the data to facilitate users' understanding. Such algorithms can range from simple rule-based filters (e.g. automatic heart failure admission based on their medical records \cite{Cox2017}), to more complicated machine learning models (e.g. \cite{Ni2019}), or visualization algorithms. 
Evaluation is therefore independent of the dashboard's user interface and also independent of the user's workflow in general.

\subsubsection{AP: Evaluation Criteria }

There are two main  perspectives for assessing a \textit{good algorithm performance}: theoretical - assessing whether an algorithm is capable given its design to generate correct outputs, and empirical, in which the algorithm's outputs are evaluated independently of its inner structure (this is sometimes referred to as a black-box test). Optionally other parameters of the algorithm may be subject to evaluation. Time and memory complexity \cite{arora2009computational} may come under scrutiny depending on available resources, data efficiency (see example in computer vision\cite{george2017generative}) may also present concerns in heavy data analysis scenarios. Essentially any evaluation metric that can be applied to the algorithm itself independent of its implementation, system localization and interaction with other architectural components such as the dashboard interface belong here (e.g. accuracy, memory and time complexity). Only seven out of 82 (8.54\%) reviewed dashboard evaluations include this scenario. 

\subsubsection{AP: Measures and Examples }

Algorithm type dictates the choice of performance measures used in evaluation. For predictive models (e.g. predicting whether a patient has high risk of a certain disease), metrics such as F-measure, Area under Receiver-Operator Characteristic (AUC), sensitivity and specificity are habitually employed (e.g. \cite{Menard2019,Cox2017}) all of which require ground truth information. Ground truth data is usually acquired through human labeling (e.g. number of patients that were diagnosed as positive by clinicians) or alternative methods in which the researchers have confidence (e.g.  medical tests). In addition, human judgements have also been used to flag mistakes directly. Ni et al. \cite{Ni2019} uses a survey to evaluate a clinical trial patient screening system, in which participants pointed out that the recommendation system did not produce consistent results, identifying the reliability issue of the automatic algorithm embedded in the system. 

We did not observe any dashboard evaluation in healthcare reporting speed, memory performance or visualization quality assessments, which are typical in visualization evaluation \cite{isenberg2013systematic}. The lack of domain specific studies might come down to the practice of reporting algorithm quality immediately when a novel algorithm is introduced (e.g.  high-dimensional data visualization \cite{bertini2011quality}), rather than in a dashboard evaluation study.

\subsubsection{AP: Evaluation Challenges}

\textbf{Fails to paint a holistic picture.} While algorithm performance represents a key evaluation marker for alerting dashboards, it is only a precursor to establishing the quality of dashboards in general.
Modern research interests have been elevated beyond assessing the quality of algorithms employed behind the scenes. For instance, judging only algorithm performance does not identify whether the usage of the dashboard effectively helps complete the intended task, or indicate whether the user perceives the interaction as a positive one. Additionally, we know from HCI that an unpleasant experience will influence whether a user continues to interact with a system/application or moves on to another \cite{Lehmann2012}. 
The dashboard's main purpose positions it as a fundamentally user facing system, for which measures of algorithm performance are essential, but far from sufficient to model the complexity of the interaction.

\subsection{System Implementation (SI)}
\label{system-2}

\textit{Does the implementation of the system fit its working environment?}

\subsubsection{SI: Evaluation Aim}
This scenario aims to \textit{assess whether the implementation of the dashboard is appropriate for the user's work environment - pays enough consideration to physical and hardware constraints specific to the primary users' work environment, and provides enough functionality or richness of data for the intended task.} This evaluation is dependent on the intended task, organisation, and social context, and independent from the user(s). Note that there is a set of literature in implementation framework of a wider range of new technologies in healthcare settings and designing comprehensive implementation strategy (e.g. \cite{May2007,Moullin2015}), which is above the scope of this work. 
Nine out of 82 dashboard (10.98\%) include this evaluation.

\subsubsection{SI: Evaluation Criteria}

The question of \textit{proper system implementation} comes down to a variety of factors and features, which we attempt to make explicit. Information systems such as dashboards exist in complex informatic ecosystems, where they serve, interact with or even incorporate modules such as databases, network modules, schedulers, data processing algorithms, frontend - backend architectures, visualization algorithms, logging systems etc. 
Therefore, as a piece of software, dashboards are placed under the same evaluation criteria as general software (e.g. \cite{ISOIEC25010}), as well as specific criteria related to 
the effectiveness of the marriage among dashboard, the intended task, and the environment it is deployed in. 
In addition, in healthcare, certain qualities take on different interpretations in view of the sensitive data these dashboard interact with. 

\subsubsection{SI: Measures and Examples}

Data for evaluating this scenario is usually obtained from domain experts in health (e.g. clinicians, policy makers) through interview, focus group discussions, observation, or questionnaires (e.g. \cite{Soh2019,Harris2018,Meijers2013,Dolan2013}) who often do not have expert knowledge of system implementation, placing responsibilities on the researchers to design appropriate instructions. For example, Harris et al. \cite{Harris2018} interviewed civil servants with multidisciplinary background in the health department, guided by Consolidated Framework for Implementation Research \cite{Damschroder2009} to identify the potential challenges that could emerge while implementing a decision-support dashboard. An alternative solution is to identify whether the participants find using the dashboard challenging or difficult through open questions, and extract a common theme (e.g. \cite{Meijers2013,Dolan2013}). The users reported implementation issues remarkably consistently - typically concerning the trust and security of data (e.g. whether the data is from a trustworthy datasource), the choice of data presentation (e.g. whether the data is in the right level of detail and appropriate format), and support (e.g. user manuals, training for adapting to the use of the dashboard).

\subsubsection{SI: Evaluation Challenges}
\textbf{Inconsistencies in user feedback.} Understanding and being familiar with the dashboard and the underlying system structure, and having the same expectations of the system (e.g. the main task outcomes, functionality) is ideally required to identify critical issues and position them explicitly with respect to system implementation. In practice this desirable state of affairs is seldom achieved among users. Thus, the feedback collected is often opaque, inconsistent, or not applicable to system implementation evaluation whatsoever. Extracting actionable issues from feedback becomes more difficult when collecting reports from a small number of users, as it leaves very small room for extracting common themes.

\subsection{Summary of the evaluation framework}
We reviewed papers that address evaluation of 82 dashboards in healthcare and extracted seven evaluation scenarios grouped into three themes:

\begin{itemize}
    \item Interaction Effectiveness
    \begin{itemize}
        \item Interaction Workflow (7/82, 8.54\%)
        \item Task Performance (43/82, 52.44\%)
        \item Behaviour Change (5/82, 6.1\%)
    \end{itemize}
    
    \item User Experience
    \begin{itemize}
        \item Perceived Engagement (49/82, 59.76\%)
        \item Potential Utility (6/82, 7.32\%)
    \end{itemize}
    \item System Efficacy 
    \begin{itemize}
        \item System Implementation (9/82, 10.98\%)
        \item Algorithm Performance (7/82, 8.54\%)
    \end{itemize}
\end{itemize}

The question of which evaluation scenario lends itself best to various types of dashboard does not admit a simple answer. In practice, the selected evaluations are restricted by external factors, such as direct access to users, or development resources. In the next section, we present a case study that exemplifies how these scenarios apply within the constraints of a real-world design process.

\section{Using the Framework: A Case Study}

To demonstrate the use of the proposed framework, we report the main visual changes of a COVID-19 dashboard throughout a 4-month iterative design period. The proposed evaluation framework is applied in validating the design choices, which is a key step in design study suggested by \cite{Munzner2009,sedlmair2012design,meyer2019criteria}. In particular, considering this is a public-facing dashboard that requires some domain-specific knowledge, we focused on three evaluation scenarios, Interaction Workflow (IW), Perceived Engagement (PE) and Potential Utility (PU) to identify the interactive and conceptual barriers for audiences with various visual literacy levels exploring the content.  We hope this provides examples on how to select evaluation scenarios based on dashboard type and sample questions to pose.
\subsection{Background }
\label{casestudy-background}

The visualization community has contributed extensively to the response to COVID-19 by creating visualizations and dashboards to illustrate the impact of the pandemic, tracking statistics updated by each country's health authority, sharing knowledge on personal health measures and potential transmission chains. Until the widespread deployment of a vaccine plan, a well-functioning contact tracing system is a vital element in reducing transmission, with five key stages (FTTIS\cite{IndependentSAGE2020}): find the population at risk (Find), test positive cases (Test), trace people who are tested positive (Trace), isolate the positive cases and their close contacts (Isolate), and support people during isolation (Support). Tracking the performance of each stage, and identifying locations or stages that urgently need improvements is essential to the system's impact as a whole. However, this represents a challenging undertaking as a dataset featuring all the five stages has yet to be released; the only alternative is represented by related datasets from only loosely connected sources with partial information on the process, that is often difficult to interpret, and laborious to curate into a uniform database.

To fill these needs, we designed a public-facing dashboard, with the aim  to combine disparate related COVID-19 datasets in England, to help researchers and policy makers locate performance data regarding each stage quickly, and public users to understand trends across the five stages. 
In June 2020, a working group was formed, consisting of 12 researchers from HCI (3 people), Epidemiology (2 people), Public Health (4 people), Communication Studies (1 person), Health Policy (1 person) and Operational Health (1 person) from the University College London and University of Leeds , who have been working in the infectious disease control field\footnote{The dashboard is released to public in Oct 2020 (https://covid.i-sense.org.uk). The full group can be found in the acknowledgements tab.}. This collaboration involved weekly group meetings, and all the sessions were chaired by either the first or the third author.

\subsection{Selecting Evaluation Scenarios}

For this COVID Response Evaluation dashboard, there are three main intended tasks:
\begin{itemize}
\item Communicating the epidemiology domain specific concepts, e.g. FTTIS, which aims to encourage exploration and concept learning for audiences with various levels of visualization literacy and domain knowledge.
\item Monitoring the key performance across FTTIS stages together with in-depth information for each stage. Presenting the rich longitudinal data (e.g. various quality indicators aggregated by locations or time) aims to provide rich resources for users to locate relatable information, and compare spatial and temporal trends which are key   epidemiological information.  
\item Serving as a data hub which collates disparate and complicated datasets. This enables motivated users who would like to trace data back to its origins either consuming related reports or working on the raw data. 
\end{itemize}

We select three evaluation scenarios, namely IW, PE and PU, to validate the dashboard design that supports these three main tasks. Considering the dashboard is public-facing and exploratory, unlike clinical decision-making or alert type dashboards which have explicit instruction on follow-up actions,  IW is selected to identify designs that are obstructive to users' natural exploring patterns performed. Study\cite{Jackson2012} has revealed that exploratory tasks with complex datasets, in this case involving domain specific concepts and spatial temporal dimensions, lead to information overload.  Therefore, PE is selected to assess users' subjective feedback (on e.g. aesthetics, usefulness and level of interest), and identify content or presentation that is overwhelming. In addition, as we aim to provide a collection of disparate datasets facilitating free-form exploration,  PU is selected to identify extra complimentary datasets and visualizations as well as detailed use cases for individual user groups (e.g. health authorities) that may be overlooked.  The weekly group discussions or interactions with the dashboard were guided with questions designed based on these three scenarios (see examples in Table \ref{tab:selectedSce}).

\begin{table}[]
\renewcommand{\arraystretch}{1.3}
    \centering
        \caption{Selected evaluation scenarios and example questions}
    \label{tab:selectedSce}
    \begin{tabular}{p{0.2\columnwidth}|p{0.7\columnwidth}}
    \hline
        \textbf{Evaluation scenarios} & \textbf{Example questions} \\
        \hline
      \multirow{6}{0.19\columnwidth}{Interaction Workflow (IW)}& Which part do you look at first on the dashboard? \\
      &Can you explain what you are trying to accomplish while using this dashboard, what steps you take and why?\\
      &Do you have any particular information in mind to look for?\\
      \hline
      \multirow{7}{0.19\columnwidth}{Perceived Engagement (PE)}&Do you find this interface aesthetically appealing?\\
      &Did you learn anything new or interesting from this dashboard?\\ 
      &What challenges and usage barriers can you see for this dashboard?\\
      &Which part of the dashboard can be reworked to improve the visual design?\\
      \hline
      \multirow{4}{0.19\columnwidth}{Potential Utility (PU)}&What would other data you would like to explore look like, despite not currently being available in this dashboard?\\
      &How do you think this dashboard can be useful and to who?\\
      \hline

    \end{tabular}

\end{table}

\begin{figure*}[t]
    \centering

\includegraphics[width=2\columnwidth]{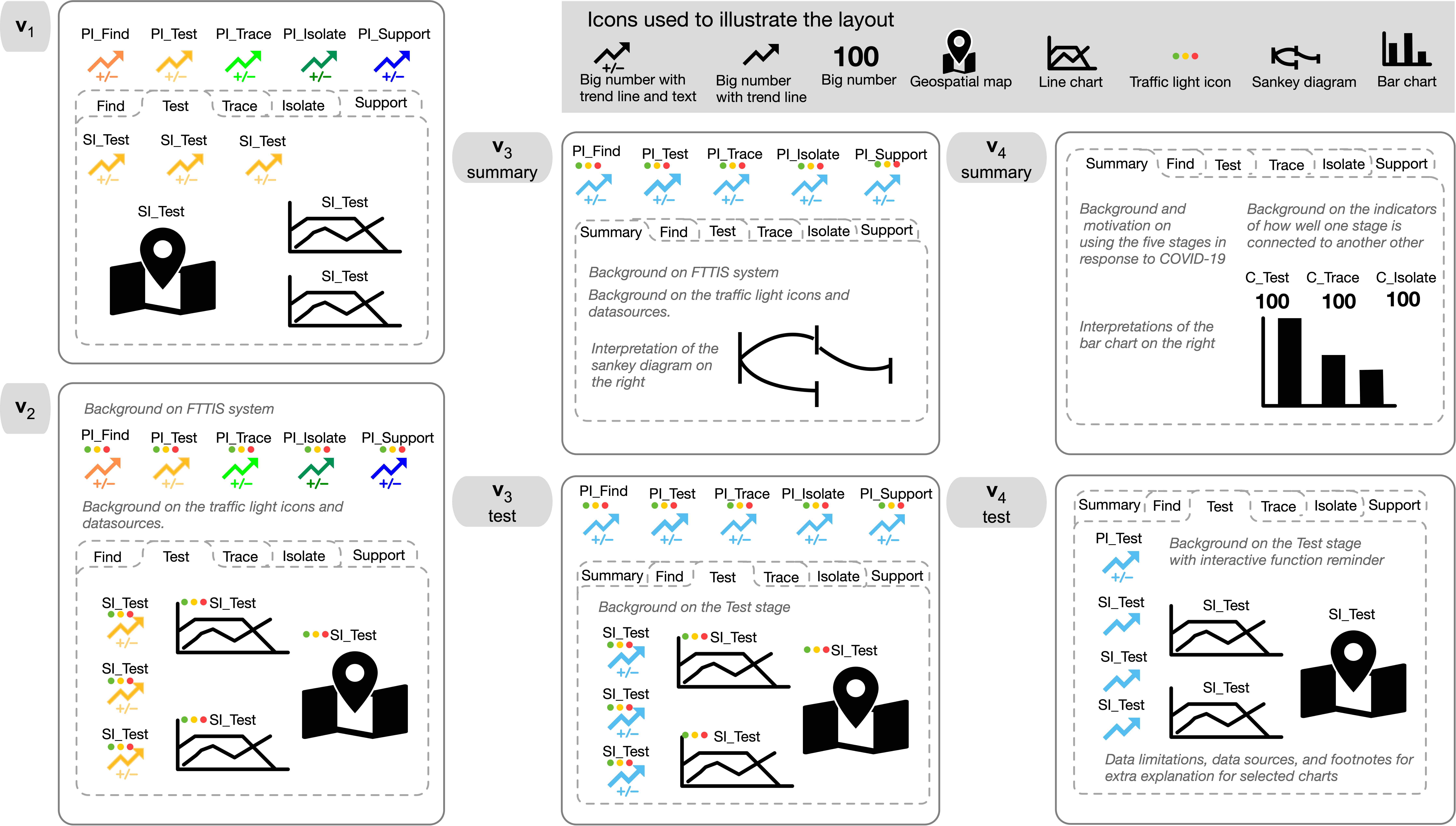}
    \caption{Illustrations of main visual changes of the COVID Response Evaluation Dashboard.  The general trend is to distill information within each context for users to consume by simplifying color coding, reducing the number of figures per page, and providing essential background information in obvious places. \emph{PI} refers to primary indicator, \emph{SI} refers to secondary indicator, and \emph{C} refers to how well the stages connect by displaying the percentage of people flowing from the initial stage to the current.}
    \label{fig:fourversions}
\end{figure*}

 Note that other evaluation scenarios, such as TP and SI, are also relevant. However, considering the public-facing and exploratory nature of the dashboard, we did not select them as the critical evaluation scenarios to validate design choices. In particular, as mentioned in section \ref{tpchallenge}, designing a task is challenging for exploration dashboards. In this case, essential tasks, locating key performance indicators of each FTTIS stage,  has been made straightforward with big number visual design (see $\textbf{v}_1$ in Fig \ref{fig:fourversions}), while focusing on more complicated tasks with a narrower scope may limit potential usage. As our dashboard is web-based using open sourced data in a wide range of settings (not limited to the workplace), we are interested in the SI criteria, which in this case means to work well with mainstream web browsers and compatible with smaller screens such as tablets and mobile phones. This criterion has been met by choosing  appropriate development tools.

\subsection{Design Changes}
\label{casestudy-process}
Fig. \ref{fig:fourversions} presents the four main versions of the dashboard, illustrated with icons. The initial design ($\textbf{v}_1$ in Fig. \ref{fig:fourversions}) contains a panel of primary indicators (PI) of each stage on the top of the page, summarising performance. Immediately below we provide a row of tabs, each providing more details of the selected stage, such as secondary performance indicators (SI),  categorical and spatial breakdowns of performance indicators. Each stage is uniquely colour coded. The colour scheme establishes visual consistency within one stage and differentiates it from the rest. Individual tabs provide more information for each stage in FTTIS, and we use the `Test' tab as  an example to demonstrate how the issues identified relate to those tabs.

\begin{figure*}[t]
    \centering

\includegraphics[width=2\columnwidth]{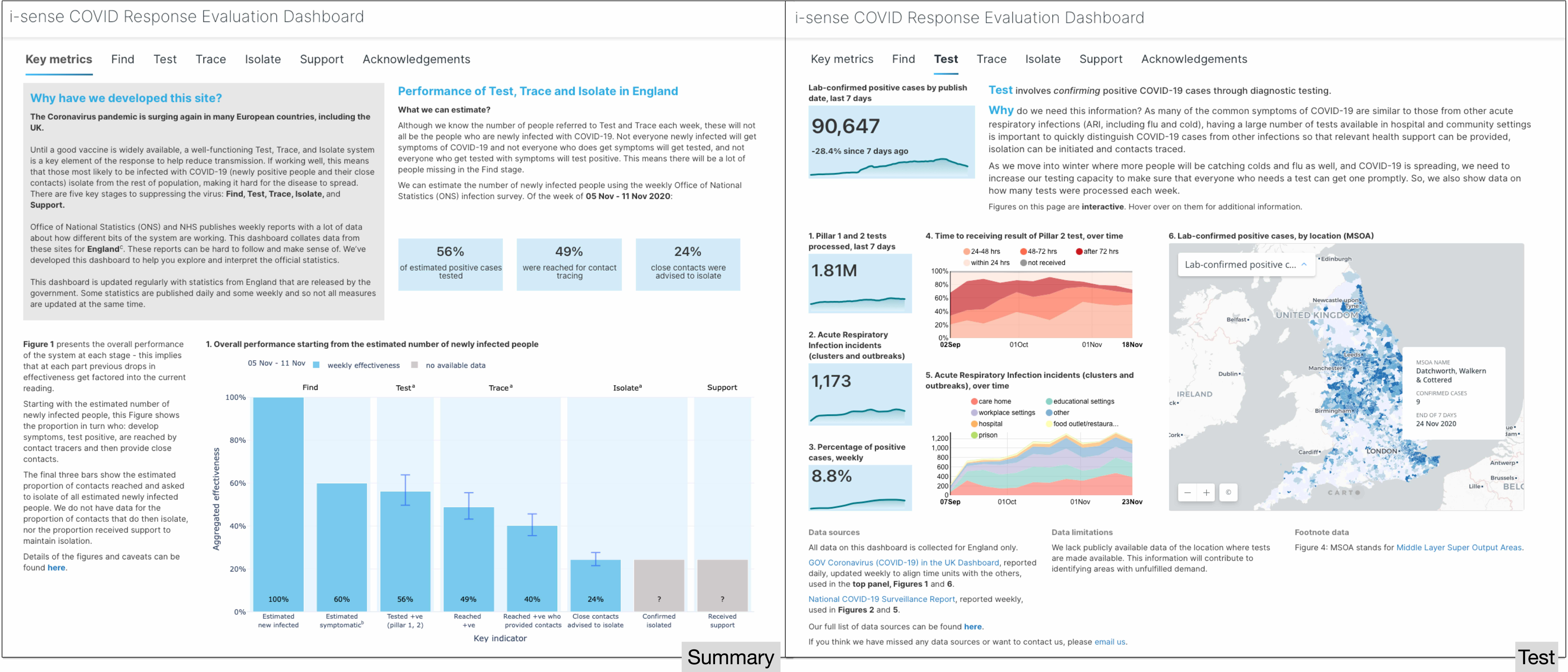}
    \caption{The COVID Response Evaluation Dashboard (date retrieved: December 1st 2020).The summary page, named \emph{Key metrics}, is shown on the left, presenting background information of the five stages and an overview on how well they connect. The `Test` page is shown on the right, presenting more related data under this heading with interactive functions for users to explore the detailed information.}
    \label{fig:finalversion}
\end{figure*}

Table \ref{tab:issues} presents the key issues raised from evaluating each design version. The final design of the dashboard is presented in Fig. \ref{fig:finalversion}. The identified issues particularly focus on introducing concepts through appropriate language and visual encoding, and reducing the risks of information overload, echoing the challenges identified in a previous general dashboard review\cite{Sarikaya2018}. Apart from identifying shared challenges and providing example solutions, we highlight the iterative need in the design process through this example, as some issues are not appropriately addressed in a single version change. In particular, as identified by PU of  $\textbf{v}_2$, the users tend to know more about the connection between individual stages. However, later, a newly created Sankey diagram was revealed to be difficult to consume while evaluating PE of $\textbf{v}_3$, which led to further change in visualisation design. Note that the dashboard changed on a weekly basis during the four months of design, and we only report major changes in the iterative process. Other changes such as adding additional data for potential utility and improving code quality for faster visual loading are not reported.

  \begin{table*}
    \centering
    \renewcommand{\arraystretch}{1.3}
        \caption{Key issues identified in each design version of the COVID Response Evaluation Dashboard, using three evaluation scenarios and solutions.}
    \label{tab:issues}
    \begin{tabular}{p{0.1\columnwidth}p{0.1\columnwidth}p{0.85\columnwidth}p{0.8\columnwidth}}
    \hline
        \textbf{Version}&\textbf{Scenario}& \textbf{Key issues identified} & \textbf{Solutions} \\
        \hline
      \multirow{12}{*}{$\textbf{v}_1$}& \multirow{8}{*}{IW} & We observed that users did not click on the lower part of the dashboard often. Follow-up feedback shows the information in the first two rows, where primary and secondary indicators are located, overloads users’ attention, preventing them from consuming the third and forth rows. & Reduce the number of rows by organising the ‘Test‘ tab into a three column structure, with secondary indicators placed vertically in a single column.\\
      \cline{3-4}

      & &We observed that users looked for information on external websites and stopped the exploration. Follow-up feedback shows the users wanted to understand the primary and secondary indicators  in detail and interpret the numbers.&Add traffic light style icons for each primary and secondary indicator visualisation. Background information on the indicators and datasource are also included in the text.\\
            \cline{2-4}
      &\multirow{4}{*}{PE}&Feedback shows that there is not enough background information on the FTTIS, leading to users experiencing difficulties interpreting the data. A commonly flagged question was “why it is important to single out these five stages?”&Background text is included in version $\textbf{v}_2$.\\
      \hline
      \multirow{8}{0.15\columnwidth}{$\textbf{v}_2$}&\multirow{4}{*}{IW}&Feedback shows that there are too many charts and text boxes appear in a single layout, without adequate amount of clues to establish the starting point of exploration.&Reduce the amount of text showing in a single layout by creating a ‘Summary‘ tab that contains the background information, traffic light icons and datasource.\\
            \cline{2-4}

      &\multirow{4}{*}{PE}&Feedback shows that the number of colours is high and some colours are similar, which makes understanding and remembering their semantic difficult.&Reduce the number of colours by changing the unique colour codes for the five individual stages to blue in version $\textbf{v}_3$.\\
            \cline{2-4}

      &\multirow{2}{*}{PU}&Feedback shows that the users want to know how consecutive stages connect with each other.&Create a sankey diagram illustrating the percentage of people flowing from the initial to the later stages.\\
      \hline
      \multirow{12}{0.15\columnwidth}{$\textbf{v}_3$}&\multirow{6}{*}{IW}&We observed that the users frequently return to the main page to look for the information related to the datasources or ask about the datasources.&Reduce users’ effort of locating the datasources by rearranging them to the bottom of each page. Create an external file that store all datasources with data caveats.\\
            \cline{3-4}

      &&We observed that some users did not attempt to interact with the charts. Feedback confirmed that the interactive features of the dashboard are not apparent to those users.&Remind the user that all charts are interactive and encourage them to mouseover them by adding text in the background area in version $\textbf{v}_4$.\\
            \cline{2-4}

      &\multirow{7}{*}{PE}&Feedback shows that the layout is overloaded, including too much text and colours&Reduce the amount of information by removing the text describing numerical trends for secondary indicators, moving the primary indicators into each tab to reduce duplication, and removing the traffic light icons.\\
            \cline{3-4}

      &&Feedback shows that users are confused with the sankey diagram.&Reduce the complexity of the figure by creating a bar chart which only presents the percentages of people captured at each stages and removing the details on why they are not captured. Explanations on how to interpret the chat are included next to the chart.\\
      \hline

    \end{tabular}

\end{table*}

\section{Discussion and Recommendations}
By reviewing dashboards in healthcare, we focused on the dashboards that have been developed in complex settings (e.g. wide spectrum of audiences, data, and use cases), and evaluated through diverse methods, ranging from surveys to randomised clinical trails lasting for years. 
We selected our tags based on empirical studies of dashboards and evaluation methods in the wider reach community, including Visualization \cite{Lam2011}, Public Health \cite{michie2012theories}, HCI and Software Engineering \cite{ISOIEC25010}. The emerged evaluation scenarios and themes therefore are relevant for dashboards in these domains. In addition, the diversity in users (e.g. variations in domain knowledge, data literacy) and use cases of healthcare dashboards provide a rich pool of practices that underpin  our framework. Although we expect some evaluation scenarios (e.g. Behaviour Change) are more common in healthcare than in other domains, our framework is general and transferable enough to be applied in a variety of fields.

Though dashboards are becoming critically important in the real world, it seems that principled discussions around dashboard evaluation are still in their infancy within the community. We observe insufficiently systematic evaluation reporting practices in dashboard evaluation studies, which is also emphasized in \cite{isenberg2013systematic} for visualization evaluation. Another observation is that not many studies integrate dashboard evaluation into dashboard design (e.g. through user-centered or iterative approaches \cite{Martinez2018,Bernard2019}). Instead of evaluation being employed as a post design method, as an opaque guarantee of performance which, by this point in time, fails to serve as a gateway to deployment, it would better serve as a constant validation of the dashboard’s purpose (ensuring that design goals align to empirical ones) and as a continuous systematic method of quality control during design stages. Moreover, principled evaluation renders the design journey more interpretable and allows other practitioners to extract mutable patterns of design in order to structure and normalize dashboard design as a whole.  Researchers would no longer need to tailor complicated measures to each dashboard, unless they were exploring novel techniques.

We therefore provide a use case of our evaluation framework, by employing selected evaluation scenarios to validate a COVID-19 dashboard design iteratively and direct design changes. Although facilitating audiences with a wide range of information literacy and avoiding information overload have been named as challenges of dashboard or visualisation design, these barriers may reflect more than one area, including context introduction, data abstraction, visual encoding, colour scheme selection, and interface arrangement to guide the user through the content naturally.  This highlights the importance of evaluating both the process of using dashboards (e.g. Interaction Workflow), and the outcome of the usage (e.g. Task Performance, Behaviour Change), through an iterative process either in the design phase or revisiting after implementation.

Based on our review and applying the evaluation framework in a real dashboard design, we make the following recommendations for future dashboard development and evaluation:

\begin{itemize}
  \setlength\itemsep{0em}
    \item \textbf{Consider diverse evaluation scenarios.}  Although more than half of the existing studies evaluate dashboards from a single scenario (e.g. 42 studies used only one of Task Performance, Perceived Engagement, or Algorithm Performance), researchers are recommended to consider all the other scenarios in order to provide a comprehensive evaluation. In addition, the interplay of two or more evaluation scenario may bring insights. For example, users who conduct different interaction patterns may reach very different perceived engagement feedback or task performance. We have not yet observed this approach applied in healthcare dashboard evaluation in this review, although the relationship between Interaction Workflow, captured through fine-grained interaction sequences and Perceived Engagement, measured through questionnaires, was identified while users were browsing an online image collection\cite{Zhuang2018}.

    \item \textbf{Learn from the past in selecting evaluation criteria and measures.} In Section \ref{sec-scenario} of this review we have detailed the evaluation criteria, measures and challenges associated with each evaluation scenario. It is recommended that all dashboard researchers make better use of these examples in designing evaluation studies. 
    
    \item \textbf{Set up the starting point.} To facilitate an intended task, a dashboard needs to capture users' visual attention effectively. Observations from the Information Seeking field \cite{Bates1989,Toms2000} reveal that browsing digital contents requires cues and visual priming to stimulate the process and to keep it in motion. It is therefore important to have clues to establish the starting point of interaction in a dashboard.  For example, in our case study (Section \ref{casestudy-process}), we distill the key take-away messages into prompts on the first page which invoke users' interests.
\end{itemize}

\section{Comparison to Existing Definitions and Frameworks} 
Our work defines dashboards as \textit{visual information systems}, and establishes an evaluation framework consisting of three themes and seven scenarios. Existing definitions (e.g. \cite{wexler2017big,few2006information,eckerson2010performance,yigitbasioglu2012review}) highlighted that dashboards are or include a visual display and their purpose is to facilitate tasks such as monitoring, learning or communication. However, due to the complex and evolving dashboard practice, these reviews hardly reach an agreement on the functionality of a dashboard (e.g. monitoring conditions that require a timely response \cite{few2006information}; identifying, exploring and communicating problems  \cite{yigitbasioglu2012review}); expanded to a wider range of purposes including support decision-making, communication and learning \cite{Sarikaya2018}). Our work builds on top of the visual genre, but does not try to specify dashboards' increasingly diverse functions.  We suggest that the dashboard is an information system, and emphasise the fact that such an interface connects with data, and provides a responsive visual display of such data to communicate with users.

There are several evaluation criteria proposed, specifically from the health community, that address the desired properties of a dashboard\cite{karami2017evaluation,west2015innovative}(e.g. user customization, security, information delivery, alerting, visual design, and integration and system connectivity, or direct impact on patients outcomes), and focus on health relevant measurements. Taking the same approach as Lam et al. \cite{Lam2011}, our evaluation framework primarily stems from the analysis of evaluation scenarios as opposed to quality measurements; moreover our framework also considers new scenarios inspired from dashboard applications in Public Health, HCI and Software Engineering. Another difference of our work form Lam et al. \cite{Lam2011} is that we focus on the quality of the interaction process and outcome from the use, rather than specify the purpose of the use (e.g. see boxed tags, including decision-making, knowledge management in Step 1 in Fig. \ref{fig:tagsprocess}). In a rapidly changing field, where novel use-cases and roles are continuously being developed such a classification is therefore advantageous, as it eschews a major element of variability in dashboard design and focuses on shared design paradigms only. In addition, as we have exemplified in individual evaluation scenarios, dashboards can provide measurable impact on groups of people outside the population of direct users (e.g. measurable improvements in patients' health). Evaluating dashboards only by the performance of direct users doing selected short-term tasks (e.g. time to complete a task) may fall short of capturing the ultimate goal of the dashboard (e.g.  evaluating dashboard used by pharmacists through number of visits of chronically ill patients to clinics \cite{Peterson2014}; other measurements in behaviour change). Overall, our work casts dashboards into the broader context of HCI and Information System Design, bringing their design, implementation, and evaluation into a wide research focus.

\section{Limitations}
A clear limitation of this work is that we reviewed empirical dashboard evaluations through related literature. However, the vast amount of evaluation practice that are not reported in literature may present a different distribution of methods. Therefore, we may have missed some important measures and criteria. Future work should involve the feedback of dashboard users and designers from the industry more closely, collecting data spanning a wider scope of the evaluation practices.  

In addition, we focused on healthcare dashboards in this study due to their wide use-cases and impact, and the theoretical diversity they offered in terms of purpose, user groups and data types. The resulting evaluation framework generalises to other fields and serves as a first attempt to a complete one. We hope to further refine this framework with examples from other fields.

\section{conclusions}
By surveying 260 papers in healthcare, we found 81 papers include at least one evaluation of dashboards. We reinterpret the evaluation scenarios for dashboards by refining and extending the work from \cite{Lam2011}. Seven evaluation scenarios, grouped into three evaluation themes, are presented with the focal question of this evaluation. We exemplify, in each scenario, what the properties of a successful dashboard must be, which empowers researchers with testing criteria. We enrich each scenario with a discussion of the practical implications extracting measurable variables of interest, and provide examples. Prominent challenges are discussed for each context. We further present the original dashboard design with iterative visual improvements based on issues identified by this evaluation framework.

The resulting framework, thus, can be used as a starting point for discussing dashboard evaluation in a more general context and further advancing the community's research interests into the problems we raise. As  aptly pointed out in \cite{Sarikaya2018}, dashboards are an ubiquitous and impactful tool for institutions and the general public alike, for professional collectives as well as for individuals. In this paper we acknowledge and identify how theories and practice from other research communities, such as Human-Computer Interaction, Information Seeking and Information Visualization, may apply to dashboards contribute to their advancement. We view our framework as a small but essential step towards consolidating this research track; in the future, we would like to see other researchers extending it beyond healthcare, re-coding our evaluation scenarios, abstracting and generalizing our results, in an effort to reach a unified theory of dashboard design, implementation and evaluation.

\ifCLASSOPTIONcaptionsoff
  \newpage
\fi



%

\bibliographystyle{IEEEtran}
\bibliography{IEEEabrv,tvcgbib}




%

\begin{IEEEbiography}[{\includegraphics[width=1in,height=1.25in,trim=60 10 70 0, clip, keepaspectratio]{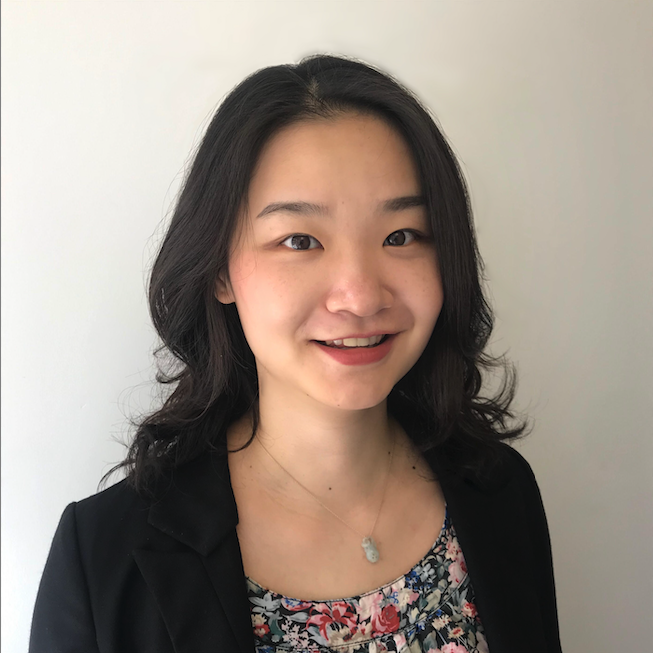}}]{Mengdie Zhuang}
Mengdie is a Lecturer in Data Science at the Information School, University of Sheffield, UK.  Her  research interests include exploratory visual analytic and how people interact with its applications  in the process of acquiring information, and how this interaction augments the user experience and influences decision-making.
\end{IEEEbiography}

\begin{IEEEbiography}[{\includegraphics[width=1in,height=1.25in,trim=250 30 250 0, clip, keepaspectratio]{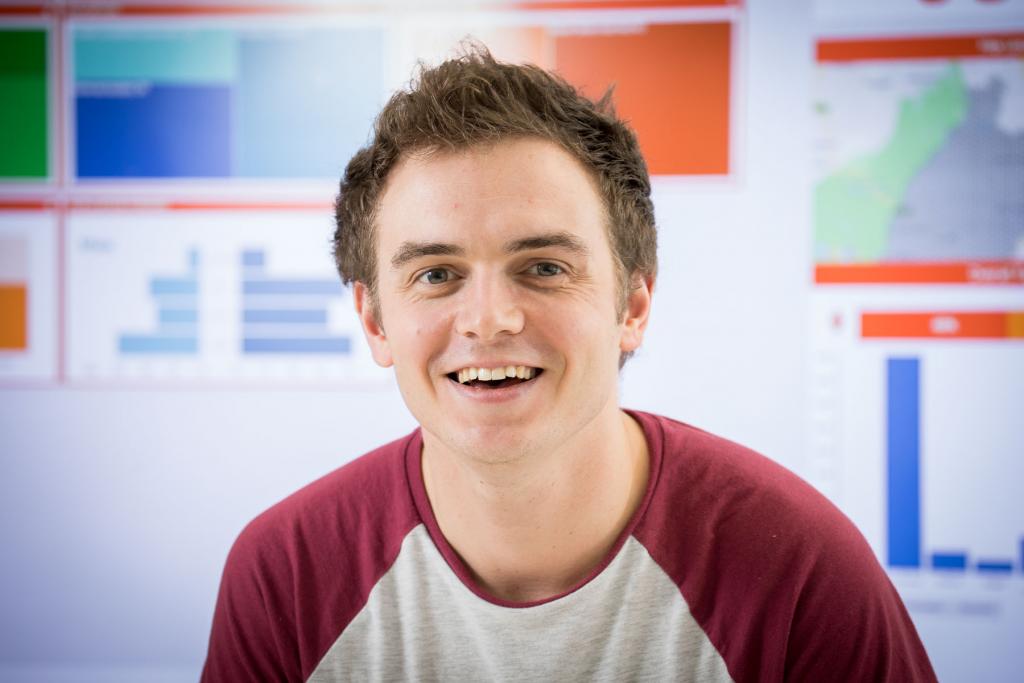}}]{David Concannon}
Dave is a PhD student at the Centre for Advanced Spatial Analysis, University College London, UK. His research focus on visualization literacy, and applying the findings in novel data visualisation development to aid decision making for the delivery of HIV care.
\end{IEEEbiography}

\begin{IEEEbiography}[{\includegraphics[width=1in,height=1.25in,trim=95 30 85 0, clip, keepaspectratio]{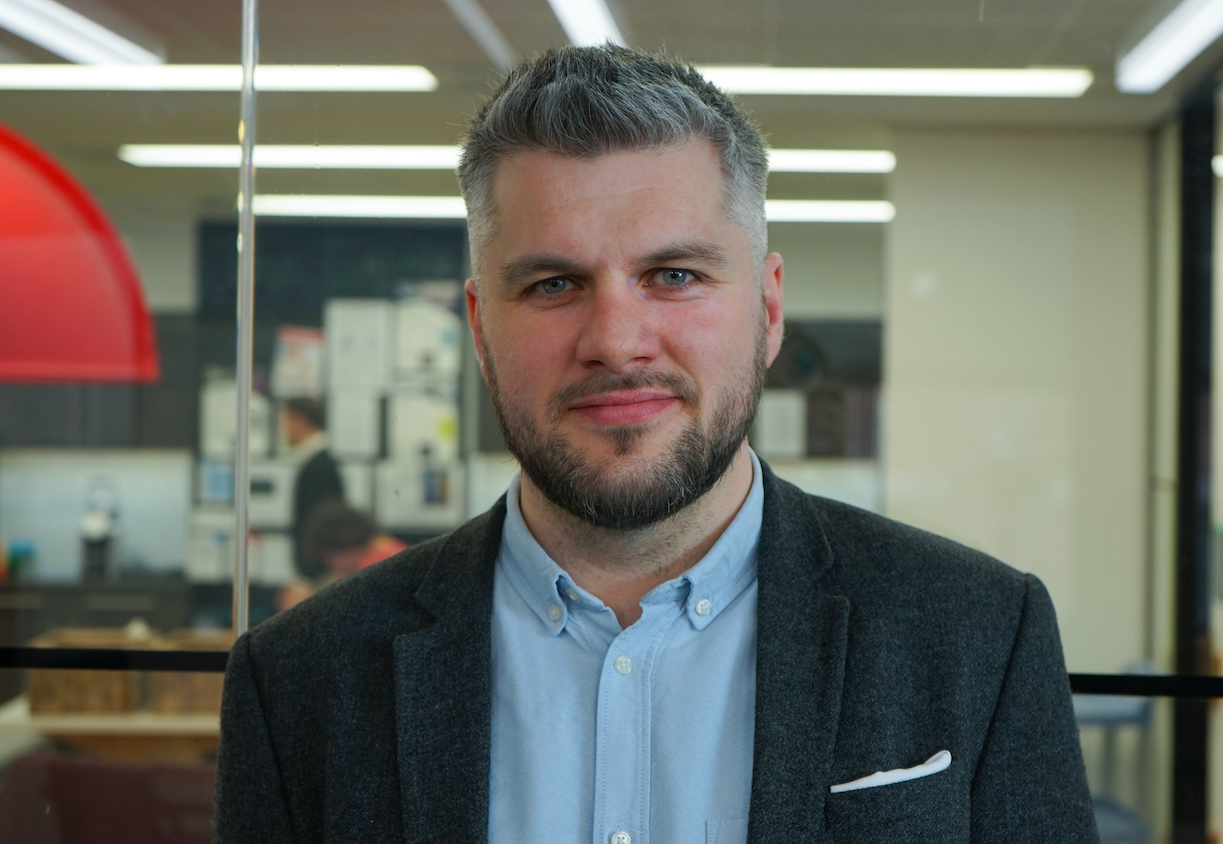}}]{Ed Manley}
Ed Manley is a Professor in Urban Analytics at the University of Leeds, UK and a Fellow of the Royal Geographical Society and Royal Society of Arts (RSA). He received an EngD from University College London.  He is author of the book ‘Agent-based Modelling and Geographical Information Science’, published by Sage in 2018. 
\end{IEEEbiography}

\vfill




\end{document}